Explanation Strategies for Image Classification in Humans vs. Current Explainable AI

Ruoxi Qi[1#], Yueyuan Zheng[1,2#], Yi Yang[2], Caleb Chen Cao[2], & Janet H. Hsiao[1,3,4]

[1] Department of Psychology
University of Hong Kong, Hong Kong SAR

[2] Huawei Research Hong Kong, Hong Kong SAR

[3] The State Key Laboratory of Brain and Cognitive Sciences
University of Hong Kong, Hong Kong SAR

[4] The Institute of Data Science
University of Hong Kong, Hong Kong SAR

# denotes co-first authorship

Address for correspondence:
Janet H. Hsiao
Department of Psychology, University of Hong Kong
Pokfulam Road, Hong Kong
Email: jhsiao@hku.hk

**Abstract**

Explainable AI (XAI) methods provide explanations of AI models, but our understanding of how they compare with human explanations remains limited. In image classification, we found that humans adopted more explorative attention strategies for explanation than the classification task itself. Two representative explanation strategies were identified through clustering: One involved focused visual scanning on foreground objects with more conceptual explanations diagnostic for inferring class labels, whereas the other involved explorative scanning with more visual explanations rated higher for effectiveness. Interestingly, XAI saliency-map explanations had the highest similarity to the explorative attention strategy in humans, and explanations highlighting discriminative features from invoking observable causality through perturbation had higher similarity to human strategies than those highlighting internal features associated with higher class score. Thus, humans differ in information and strategy use for explanations, and XAI methods that highlight features informing observable causality match better with human explanations, potentially more accessible to users.

## Introduction

To ensure good use of AI to humans, researchers have long recognized the importance of explanation to enhance human-AI interaction, including the development of Expert Systems in 1980s and Knowledge-Based Tutors in 1990s and early 2000s (Mueller et al., 2019). Around the mid-2010s, a new generation of explainable AI (XAI) emerged due to the advance of deep learning methods, whose decision-making processes are often obscured to both users and developers. As compared with previous explanation solutions, these XAI methods use better visualization techniques (Goyal et al., 2016) or directly make the classifiers themselves more explainable (Akata, 2013). However, similar to previous efforts, they remain focusing on using more AI to explain AI without much consideration of users' mental processes (Hoffman et al., 2018; Hsiao, Ngai, et al., 2021). This differed significantly from how humans provide explanations. For example, in visual explanations, human explanations typically involve directing attention to relevant details following a sequence of visual reasoning processes, in contrast to XAI methods that simply highlight features used by AI classifiers without temporal information. Human explanations also often consider explainees' prior knowledge and qualitative reasoning styles, which are typically missing in current XAI methods (Kaufman & Kirsh, 2022). Indeed, effective human explanations often involve causal reasoning based on observed regularities in the world (Einhorn & Hogarth, 1986; Maxwell, 2004; Holzinger et al., 2019; Bender, 2020). When providing explanations to others, people use more observable behavior (Malle & Knobe, 1997). They also prefer explanations that invoke causality (Zemla et al., 2017). In particular, rather than listing all possible causes of an event in an explanation, people tend to provide contrastive explanations that focus on why the current event occurs instead of other non-occurring events (Chin-Parker & Cantelon, 2017; Miller, 2021; van Fraassen, 1980). Knowledge about how humans give explanations provides important insights into ways to make explanations from XAI more accessible to humans.

Despite these initial efforts, our current understanding of how humans provide explanations on tasks that are commonly performed by AI remains very limited, especially for those involving making decisions based on complex perceptual processes that are often automatic and unconscious in humans such as image classification. Image classification has been a heated topic in computer vision, and the advance of deep learning methods in recent years has significantly increased automated image classification accuracy (Rawat & Wang, 2017). A common XAI method for image classification has been using saliency maps that highlight regions of the input image according to their importance to the AI model's output (Li et al., 2021). Two major saliency-map based approaches are perturbation-based and backpropagation-based methods. Perturbation-based methods, such as RISE (Petsiuk et al., 2018), perturb the input image and place more weights on the pixels that affect the output class probability relative to other classes when occluded. In contrast, backpropagation-based methods, such as GradCAM (Selvaraju et al., 2020), calculate the gradient of the score for the target class in a particular layer as the class relevance of each pixel. These saliency maps have often been compared with human attention maps under the assumption that humans attend to features important to their judgements during image classification (Hwu et al., 2021; Lai et al., 2020), and thus human attention may provide a good benchmark for their plausibility (e.g., Mohseni et al., 2021; Yang et al., 2022; Karim et al., 2022; Lanfredi et al., 2021).

Note however that human attention when viewing an image has been shown to be task-driven and thus may differ significantly when the task demand changes (e.g., Kanan et al., 2015; Hsiao, An et al., 2021). On one hand, saliency-map based XAI highlights image regions that contribute to the classifier output, and thus should be compared with human attention when performing image classification tasks (Lai et al., 2020). On the other hand, the purpose of the saliency maps is to provide explanations, and thus they may be better compared with human attention when they explain image classification (Yang et al., 2022). It remains unclear how human attention differs between image classification and explanation tasks. Providing explanations on how to perform a task involves metacognitive skills to evaluate thought processes through self-awareness (e.g., Jiang et al., 2016; Balcikanli, 2011), and thus explanation strategies for the task may differ from the strategies for performing the task itself. Also, during image classification, humans only need to attend to sufficient information for making a decision (Hsiao & Cottrell, 2008; Smith & Ratcliff, 2004), whereas during explanation, they may attend to all relevant information to provide a comprehensive explanation (Gelman et al., 1998). Thus, human attention during image classification and explanation may differ significantly. Understanding whether saliency maps generated using the current XAI methods are better matched with human attention during image classification or explanation will provide important insights on what information these XAI salience maps reflect and how human users should interpret them.

Another factor to consider is the substantial individual differences in human attention during cognitive tasks as demonstrated through eye tracking studies (e.g., Chuk et al., 2014; Chuk, Crookes, et al., 2017; Hsiao, Chan, et al., 2021; Hsiao, Lan, et al., 2021; Peterson & Eckstein,

2013), and these individual differences are often associated with differences in task performance and cognitive abilities (e.g., An & Hsiao, 2021; Hsiao, Lan et al., 2021). As individuals can differ significantly in both cognitive and metacognitive abilities across domains (e.g., Rouault et al., 2018), substantial individual differences in explanation strategies are expected. It remains unclear how individual differences in explanation strategy are compared with current XAI methods. More specifically, human object category representations involve both visual and abstract conceptual features (Martin et al., 2018), both of which can be used in explanations. Thus, individuals may differ in their reliance on visual or conceptual information when providing explanations, and XAI saliency maps may match better with attention strategies associated with more use of visual information, which may also be more effective for early learners without prior knowledge (Fisher & Sloutsky, 2005). Also, since human prefers explanations that involve causality and contrastive explanations (Zemla et al., 2017; Miller, 2021), saliency maps generated through perturbation methods, which use observable causality between input perturbation and consequent change in output class probability, may show a better match with human attention during explanation than gradient-based methods. The comparisons between different human and XAI explanation strategies will provide important insights on the explanation processes of both humans and XAI methods and ways to enhance XAI explanations to facilitate human-AI interaction.

To test these hypotheses, here we examined human attention strategies in image classification and in explaining image classification, whether individual differences in these strategies are associated with classification and explanation performance and visual vs. conceptual information use in the explanations, and how human attention strategies are compared with the saliency maps generated by current XAI methods. We used eye tracking to directly measure human attention, in contrast to indirect measures such as the annotation or pointing approaches that are typically used in previous studies (e.g., Mohseni et al., 2021; Gelman et al., 1998). To quantify individual differences in visual scanning behavior during the tasks and to discover representative attention strategies in humans, we adopted a machine-learning-model based approach, Eye Movement analysis with Hidden Markov Models (EMHMM; Chuk et al., 2014) with co-clustering (Hsiao, Lan, et al., 2021). In this approach, an individual's eye movement behavior in viewing a stimulus is summarized in a hidden Markov model (HMM) in terms of person-specific regions of interest (ROIs) and transition probabilities among the ROIs. The co-clustering algorithm is then used to discover participant groups where group members adopt similar strategies to one another across stimuli, with each group forming a representative attention strategy. Similarities among individual strategies then can be quantitatively assessed using their data log-likelihoods given the representative strategy models. Consistency of a strategy can be assessed using entropy of the HMM (Cover & Thomas, 2006; higher entropy indicates lower consistency). Thus, adopting this approach allows us to take both spatial (ROI choice) and temporal information (the order of the ROIs visited) into account when quantifying individual differences in attention strategies. This approach has been applied to a variety of research fields and led to novel findings not discoverable by traditional methods (e.g., summary statistics of eye movement in predefined ROIs or fixation heatmaps; Barton et al., 2006; Caldara & Miellet, 2011), including psychology (e.g., An & Hsiao, 2021; Hsiao, Chan et al., 2021; Hsiao, An et al., 2022), mental health (e.g., Zhang et al., 2019; Chan et al., 2020), and education (Zheng et al., 2022).

We compared human attention strategies with saliency maps generated by a representative perturbation-based method, RISE (Petsiuk et al., 2018), and a representative backpropagation-based method, GradCAM (Selvaraju et al., 2020) for an image classification AI model ResNet50, which has excellent classification performance (He et al., 2016). Since RISE can be affected by the pixel distribution of the random masks used to generate saliency maps, we included a pixel coverage bias (PCB) corrected version of RISE (Xie et al., 2022). We hypothesized that human attention strategies when explaining image classification results would cover more relevant features than those during image classification itself, and individuals differ in their reliance on visual or conceptual information. Human attention maps from those who rely more on visual information would show higher similarity to XAI saliency maps, and perturbation-based explanations such as RISE, which highlights discriminative features by invoking causality through perturbation, may show higher similarity to these human attention maps.

## Results

### Image Classification Task

Participants named the category of a presented image as quickly as possible with eye tracking. EMHMM with co-clustering resulted in two representative pattern groups: Explorative (Group A, with larger ROIs) vs. Focused (Group B, with smaller, focused ROIs) Pattern Groups (Figure 1A). The two groups differed significantly based on KL divergence estimation (Chuk et al., 2014): $F$(1,

59) = 187.15, $p < .001$, $\eta_p^2 = .76$, 90% CI = [.66, .81][1]. Explorative participants had a significantly larger number of fixations than focused participants, $t(59) = 3.49$, $p < .001$, $d = 0.90$, 95% CI [0.34, 1.44]. They did not differ in average fixation duration, $t(59) = -.41$, $p = .685$, $d = -0.11$, 95% CI [−0.61, 0.40], or in eye movement entropy (consistency), $t(58.10) = -.42$, $p = .674$, $d = -0.11$, 95% CI [−0.61, 0.40].

## A. Classification task

Group A: Explorative Pattern Group (N = 34)     Group B: Focused Pattern Group (N = 27)

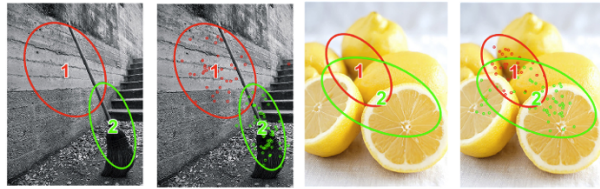 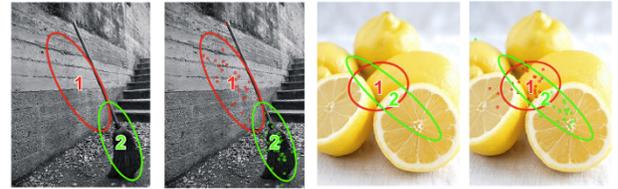

| Group A | To R | To G | Group A | To R | To G | Group B | To R | To G | Group B | To R | To G |
|---|---|---|---|---|---|---|---|---|---|---|---|
| Priors | .83 | .17 | Priors | .81 | .19 | Priors | .92 | .08 | Priors | .71 | .29 |
| From Red | .13 | .87 | From Red | .06 | .94 | From Red | .16 | .84 | From Red | .04 | .96 |
| From Green | .25 | .75 | From Green | .07 | .93 | From Green | .36 | .64 | From Green | .10 | .90 |

## B. Explanation task

Group A: Explorative Pattern Group (N = 47)     Group B: Focused Pattern Group (N = 15)

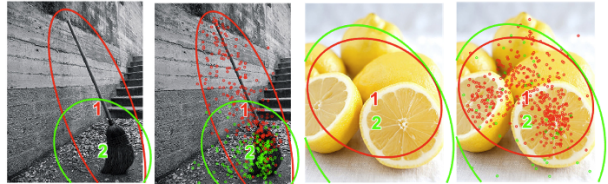 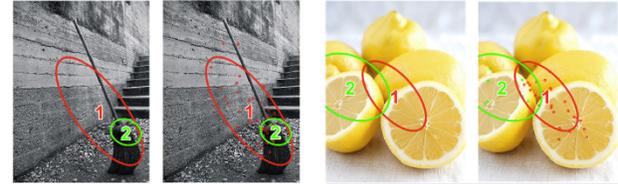

| Group A | To R | To G | Group A | To R | To G | Group B | To R | To G | Group B | To R | To G |
|---|---|---|---|---|---|---|---|---|---|---|---|
| Priors | .89 | .11 | Priors | .88 | .12 | Priors | .91 | .09 | Priors | .78 | .22 |
| From Red | .96 | .04 | From Red | 1.0 | .00 | From Red | .96 | .04 | From Red | .96 | .04 |
| From Green | .10 | .90 | From Green | .00 | 1.0 | From Green | .08 | .92 | From Green | .08 | .92 |

## C. Combining classification and explanation tasks

Group A: Explorative Pattern Group (N = 60)     Group B: Focused Pattern Group (N = 63)

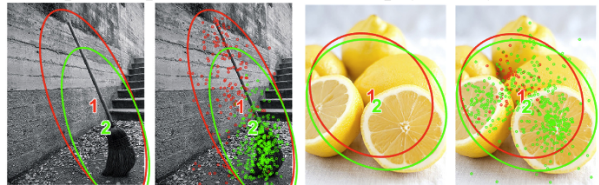 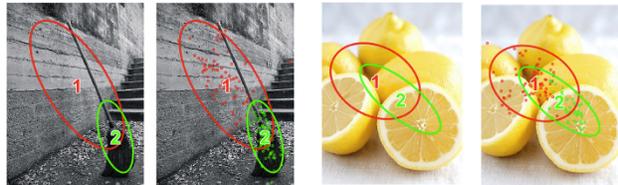

| Group A | To R | To G | Group A | To R | To G | Group B | To R | To G | Group B | To R | To G |
|---|---|---|---|---|---|---|---|---|---|---|---|
| Priors | 1.0 | .00 | Priors | .99 | .01 | Priors | .94 | .06 | Priors | .91 | .09 |
| From Red | .42 | .58 | From Red | .24 | .76 | From Red | .32 | .68 | From Red | .25 | .75 |
| From Green | .47 | .53 | From Green | .08 | .92 | From Green | .38 | .62 | From Green | .39 | .61 |

---

[1] 90% CI instead of 95% CI is reported for F-tests since F-tests are one-sided (Steiger, 2004).

Figure 1. Examples of EMHMM co-clustering results of broom and lemon images for (A) the classification task only, (B) the explanation task only, and (C) two tasks together. Ellipses show ROIs as 2-D Gaussian emissions. The table shows transition probabilities among the ROIs, and priors show the probabilities that a fixation sequence starts from the ellipse. In each pattern, the image on the right shows raw fixations and their ROI assignment.

English proficiency, as measured by LexTALE, was correlated with both classification accuracy, $r(60) = .28$, $p = .026$, and RT, $r(55) = -.36$, $p = .007$. ANCOVA analyses examining the effect of eye movement pattern group on accuracy and RT with LexTALE as a covariate showed a significant effect of group in RT, $F(1, 52) = 6.66$, $p = .013$, $\eta_p^2 = .11$, 90% CI = [.01, .25], but not in accuracy. Participants in the focused group had shorter RT, $M_{adj} = 945$, $SE_{adj} = 19.3$, than those in the explorative group, $M_{adj} = 1015$, $SE_{adj} = 18.9$, $t(52) = 2.58$, $p = .013$, $d = 0.70$, 95% CI [0.14, 1.25]. We then quantified each participants' eye movement pattern along the dimension contrasting Explorative and Focused Pattern Groups using EF scale, defined as $(E - F) / (E + F)$, where E and F referred to data loglikelihood given Explorative and Focused Pattern Group models respectively. Hierarchical multiple regression predicting classification RT showed that, at stage one, cognitive ability measures and LexTALE jointly explained 30.6% of the variance; the regression model was not significant, $F(13, 40) = 1.36$, $p = .223$. Adding EF scale accounted for an additional 6.4% of the variance and this change was marginally significant, $F(1, 39) = 3.96$, $p = .054$. Together the results suggested that the focused attention strategy was associated with shorter image classification RT.

**Explanation Task**

Participants typed explanations in a textbox about why a presented image belonged to a presented class label with eye tracking, and eye movement in the image area was analyzed using EMHMM with co-clustering. The results similarly showed Explorative and Focused Pattern Groups (Fig. 1B) that differed significantly: $F(1, 60) = 122.77$, $p < .001$, $\eta_p^2 = .67$, 90% CI = [.55, .74]. Explorative participants had significantly more fixations per trial, $t(57.78) = 6.57$, $p < .001$, $d = 1.53$, 95% CI [1.26, 2.62], longer average fixation duration, $t(60) = 2.34$, $p = .022$, $d = 0.70$, 95% CI [0.09, 1.29], and higher eye movement entropy (lower consistency), $t(37.01) = 8.34$, $p < .001$, $d = 2.20$, 95% CI [1.66, 3.27] than focused participants. In addition, explorative participants had more fixations on the image region, $t(60) = 4.02$, $p < .001$, $d = 1.19$, 95% CI [0.56, 1.82], but fewer fixations on the textbox region, $t(60) = 3.38$, $p = .001$, $d = 1.00$, 95% CI [0.38, 1.61], than focused participants.

We evaluated participants' explanations in terms of (1) effectiveness for teaching image classification to someone without prior knowledge of image classes as rated by two computer vision experts, and (2) diagnosticity as measured by naïve observers' performance in inferring the image class from the explanation. Language proficiency as measured in LexTALE was significantly correlated with effectiveness, $r(60) = .28$, $p = .029$, but not with diagnosticity, $r(60) = .10$, $p = .430$. ANCOVA analyses examining the effect of eye movement pattern group on these two explanation performance measures with LexTale controlled showed significant differences in effectiveness, $F(1, 59) = 12.71$, $p < .001$, $\eta_p^2 = .18$, 90% CI [.05, .31], and diagnosticity, $F(1, 59) = 16.74$, $p < .001$, $\eta_p^2 = .22$, 90% CI [.08, .36]: explorative participants' explanations were rated higher for effectiveness, $M_{adj} = 4.4$, $SE_{adj} = 0.09$, than focused participants, $M_{adj} = 3.7$, $SE_{adj} = 0.17$, $t(59) = 3.57$, $p < .001$, $d = 1.06$, 95% CI [0.44, 1.69]. In contrast, focused participants' explanations had higher diagnosticity, $M_{adj} = .63$, $SE_{adj} = .02$, than explorative participants, $M_{adj} = .53$, $SE_{adj} = .01$, $t(59) = 4.09$, $p < .001$, $d = 1.22$, 95% CI [0.58, 1.85]. Consistent with these findings, EF scale was positively correlated with effectiveness, $r(60) = .45$, $p < .001$ (Figure 2a), and negatively correlated with diagnosticity, $r(60) = -.42$, $p < .001$ (Figure 2b).

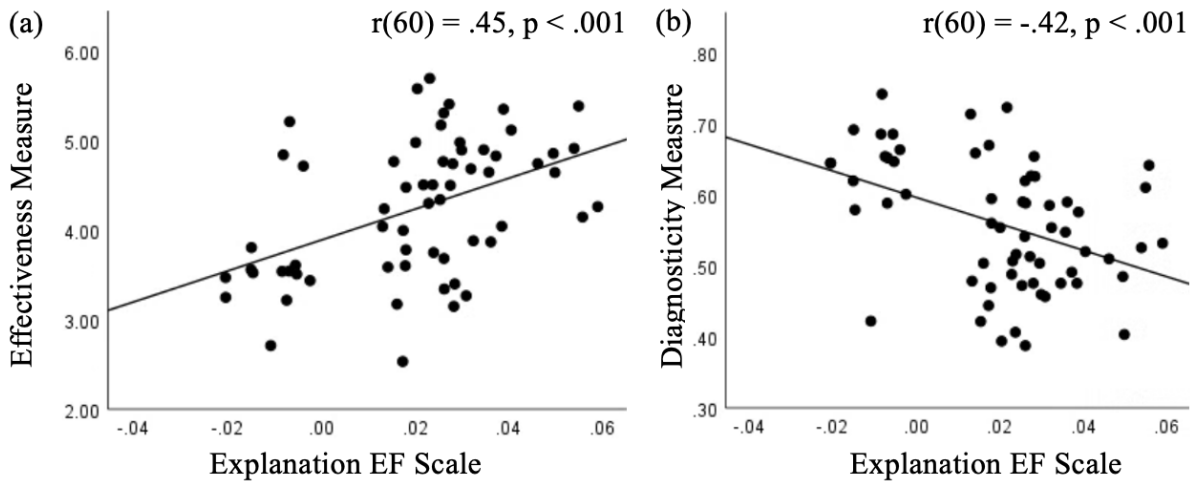

Figure 2. Correlation between EF scale in the explanation task and explanation performance as evaluated by (a) effectiveness, and (b) diagnosticity.

Hierarchical multiple regression analyses predicting explanation effectiveness showed that, at stage one, the cognitive ability measures and LexTALE contributed significantly to the regression model, $\Delta R^2 = 39.2\%$, $F(13, 47) = 2.33$, $p = .017$, and at stage two EF scale significantly explained additional variations, $\Delta R^2 = 9.9\%$, $F(1,46) = 8.98$, $p = .004$. For predicting diagnosticity, cognitive ability measures and LexTALE did not contribute significantly to the regression model at stage one, $\Delta R^2 = 28.2\%$, $F(13, 47) = 1.42$, $p = .186$, while EF scale significantly explained additional variance at stage two, $\Delta R^2 = 14.0\%$, $F(1, 46) = 11.16$, $p = .002$. Thus, after taking English proficiency and cognitive abilities into account, the explorative strategy was associated with more effective explanations for teaching image classification, whereas the focused strategy was associated with more diagnostic explanations for inferring class labels.

We also quantified the text characteristics of the participants' explanations in terms of (1) reliance on visual information as measured by visual strength of the words in the explanation according to human ratings from the Lancaster Sensorimotor Norms (Lynott et al., 2020), and (2) reliance on conceptual information as measured by WordNet similarity (Miller, 1995) of the words in the explanation to the class label. Visual strength was positively correlated with EF scale, $r(60) = .44$, $p < .001$, and effectiveness, $r(60) = 0.50$, $p < .001$, but negatively correlated with diagnosticity, $r(60) = -.36$, $p = .004$. In contrast, WordNet similarity was negatively correlated with EF scale, $r(60) = -.29$, $p = .023$, and effectiveness, $r(60) = -.42$, $p < .001$, but positively correlated with diagnosticity, $r(60) = .31$, $p = .013$. These results suggested that explorative strategies were associated with more use of visual information and less use of conceptual information. In addition, explanations with more visual information tended to be rated higher for effectiveness, while those with more conceptual information tended to be more diagnostic.

**Comparison of the Two Tasks**

No significant correlation was found between participants' EF scale of the image classification task and the explanation task, $r(59) = .16$, $p = .210$, suggesting that participants did not use consistent attention strategies across the two tasks. To compare attention strategies in the two tasks directly, we used EMHMM with co-clustering on participants' eye movement patterns in both tasks together. The results showed similar Explorative and Focused Pattern Groups (Figure 1C) that differed significantly: $F(1, 121) = 294.30$, $p < .001$, $\eta_p^2 = .71$, 90% CI = [.64, .76]. Explorative Pattern Group had higher entropy (lower consistency) than Focused Pattern Group, $t(78.47) =$

13.95, $p < .001$, $d = 2.54$, 95% CI [1.98, 3.04]. Interestingly, most participants' eye movement patterns in the image classification task were classified into the Focused Pattern Group whereas most participants' eye movement patterns in the explanation task were classified into the Explorative Pattern Group, $\chi^2(1, N = 123) = 61.6$, $p < .001$ (Table 1).

| Group | Task | | Total |
|---|---|---|---|
| | Classification | Explanation | |
| Explorative | 8 | 52 | 60 |
| Focused | 53 | 10 | 63 |
| Total | 61 | 62 | 123 |

Table 1. Number of participants' eye movement patterns during the image classification and explanation tasks clustered into the Explorative or Focused Pattern Groups.

Repeated-measures ANOVA was used to examine the effect of task (classification vs. explanation) and image type (natural vs. artificial) on participants' eye movement pattern as measured in EF scale. We found a main effect of task, $F(1, 60) = 144.81$, $p < .001$, $\eta_p^2 = .71$, 90% CI [.60, .77], where participants' eye movement patterns were more explorative during explanation than classification; and a main effect of image type, $F(1, 60) = 15.87$, $p < .001$, $\eta_p^2 = .21$, 90% CI [.07, .35], where participants' eye movement patterns were more explorative when viewing natural images than artificial ones. There was an interaction between task and image type, $F(1, 60) = 21.24$, $p < .001$, $\eta_p^2 = .26$, 90% CI [.11, 40]: participants' eye-movement patterns were more explorative when viewing natural images than artificial images in the explanation task, $t(61) = 5.20$, $p < .001$, $d = 0.66$, 95% CI [0.38, 0.93], but not in the classification task, $t(60) = -0.39$, $p = .698$, $d = -0.05$, 95% CI [−0.30, 0.20].

**Comparison with XAI Saliency Maps**

To compare with XAI saliency maps, we generated human attention maps by applying a Gaussian filter with a standard deviation equivalent to 0.5° of visual angle on each fixation (Figure 3). Cosine similarity and KL divergence were used to measure the similarity between an XAI saliency map and a human attention map. We conducted by-item four-way ANOVA to examine the effect of task (classification vs. explanation), attention strategy (explorative vs. focused), XAI method

(RISE vs. PCB corrected RISE vs. GradCAM), and image type (natural vs. artificial) on the similarity measures. Results were summarized in Table 2. As shown in Table 2, a three-way interaction between task, strategy, and XAI method was found in both similarity measures. When we split the data by XAI methods, main effects of task and strategy, and an interaction between task and strategy were consistently found across the XAI methods ($p$s < .001). XAI saliency maps had higher similarity to human attention maps during the explanation task, particularly for individuals adopting the explorative strategy, which was associated with more fixations on the image area, suggesting a higher reliance on visual information.

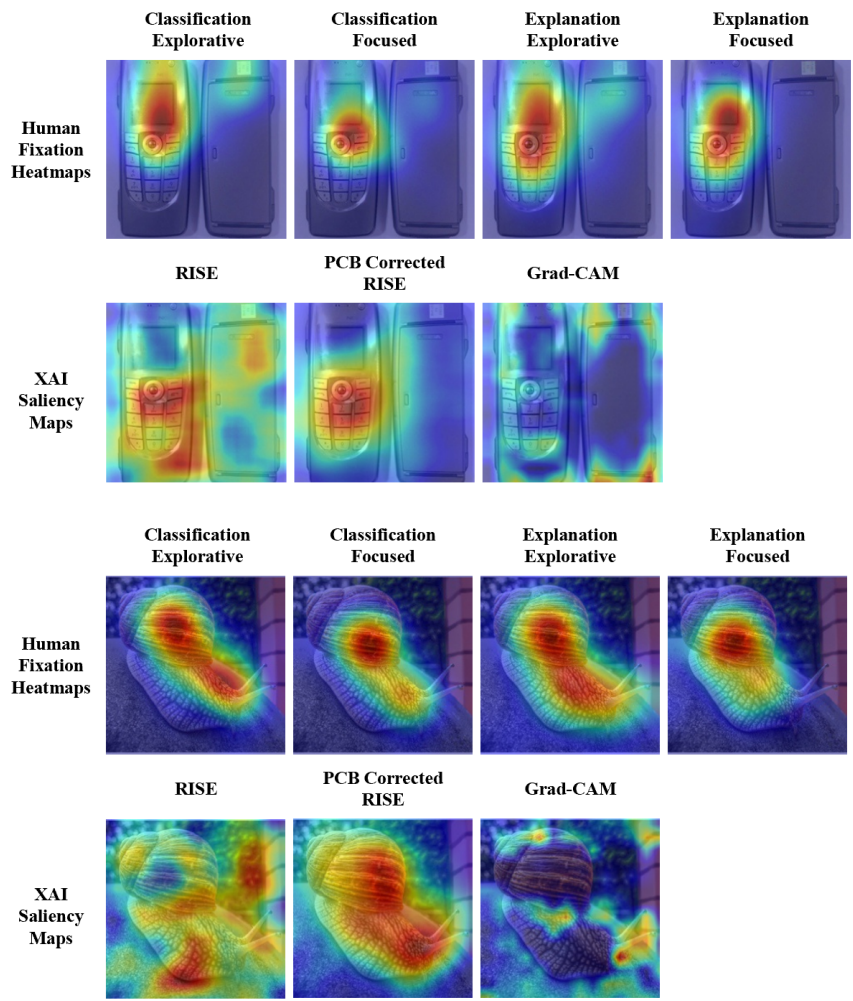

Figure 3. Example human attention maps and XAI saliency maps, with one image from an artificial category (cellphone) and one image from a natural category (snail).

| Effect | Cosine Similarity | | | KL Divergence | | |
|---|---|---|---|---|---|---|
| | $F$ | $p$ | $\eta_p^2$ | $F$ | $p$ | $\eta_p^2$ |
| Task[†] | 242.90 | < .001*** | .61 | 155.59 | < .001*** | .50 |
| Strategy[†] | 300.97 | < .001*** | .66 | 342.98 | < .001*** | .58 |
| XAI Method[†] | 270.88 | < .001*** | .63 | 177.82 | < .001*** | .53 |
| Image Type | 0.68 | .411 | .00 | 5.22 | .024* | .03 |
| Task × Strategy[†] | 240.22 | < .001*** | .60 | 120.83 | < .001*** | .43 |
| Task × XAI Method[†] | 38.27 | < .001*** | .20 | 20.96 | < .001*** | .12 |
| Strategy × XAI Method | 3.31 | .065 | .02 | 66.22 | < .001*** | .30 |
| Task × Image Type | 1.61 | .207 | .01 | 16.71 | < .001*** | .10 |
| Strategy × Image Type | 11.26 | < .001*** | .07 | 0.41 | .523 | .00 |
| XAI Method × Image Type | 2.70 | .092 | .02 | 1.36 | .254 | .01 |
| Task × Strategy × XAI Method[†] | 75.30 | < .001*** | .32 | 11.02 | < .001*** | .07 |
| Task × Strategy × Image Type | 4.86 | .029* | .03 | 2.83 | .094 | .02 |
| Task × XAI Method × Image Type | 1.20 | .284 | .01 | 4.47 | .029* | .03 |
| Strategy × XAI Method × Image Type | 2.19 | .138 | .01 | 0.42 | .589 | .00 |
| Task × Strategy × XAI Method × Image Type | 0.01 | .950 | .00 | 1.80 | .180 | .01 |

Table 2. Results of the task × eye movement strategy × XAI method × image type ANOVA on cosine similarity and KL divergence ([†]consistent across the two metrics; *p < .05, **p < .01, ***p < .001). The Greenhouse-Geisser correction was used for all of the effects that involved XAI methods due to violations of the sphericity assumption.

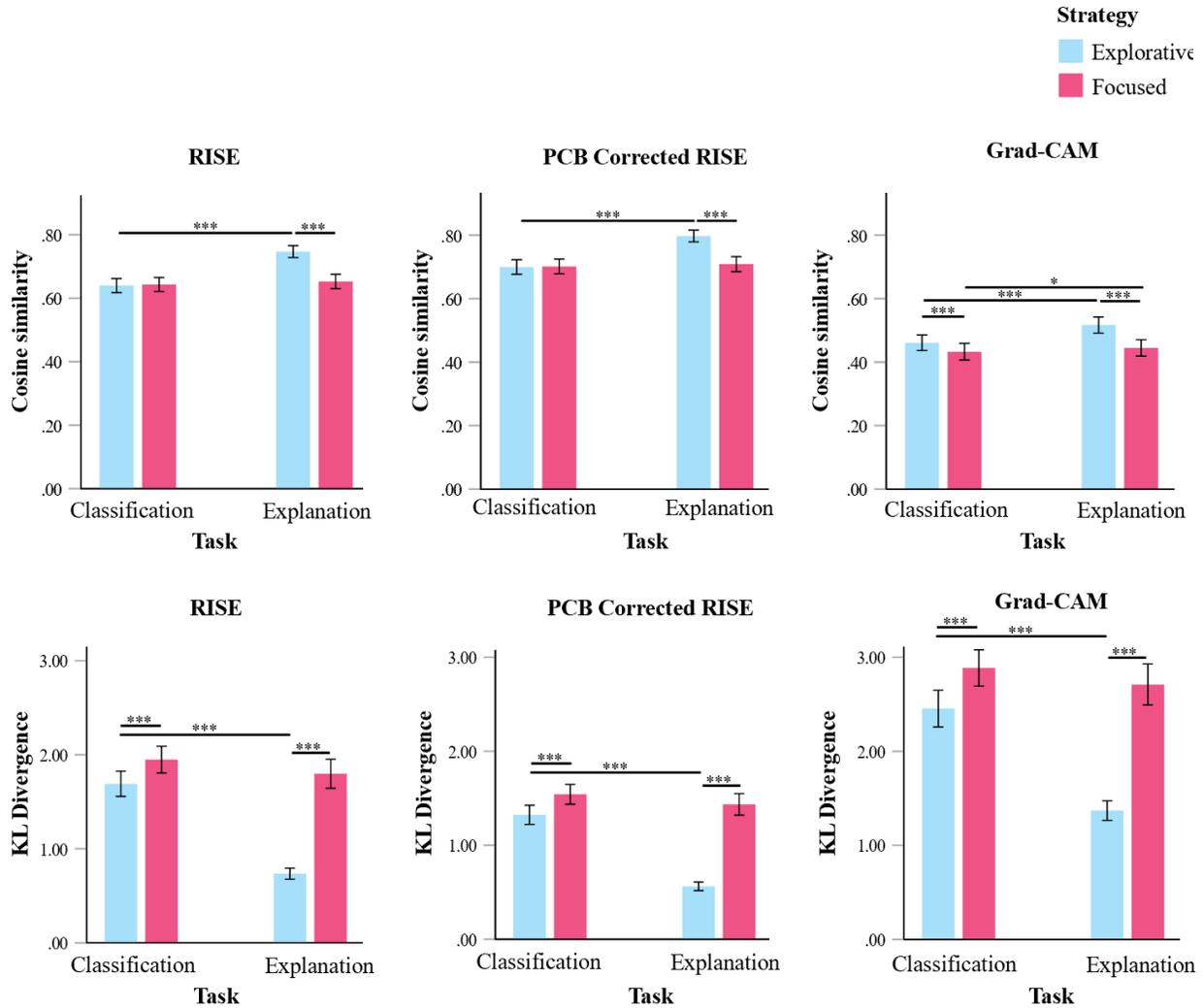

Figure 4. Difference in cosine similarity (row above) and KL divergence (row below) between the two strategies for the two tasks and each of the three XAI methods (error bars: 95% CI; *p < .05, **p < .01, ***p < .001). Note that greater similarity is indicated by higher cosine similarity and lower KL divergence.

We then split the data by task and strategy to examine the effect of XAI method. In all combinations of task and strategy, we found a main effect of XAI method in both similarity measures ($ps < .001$): Saliency maps from PCB corrected RISE had the highest similarity to human attention maps, followed by RISE, and then by GradCAM (Figure 5), suggesting that saliency maps generated by perturbation-based XAI methods have higher similarity to human attention maps than those from gradient-based methods.

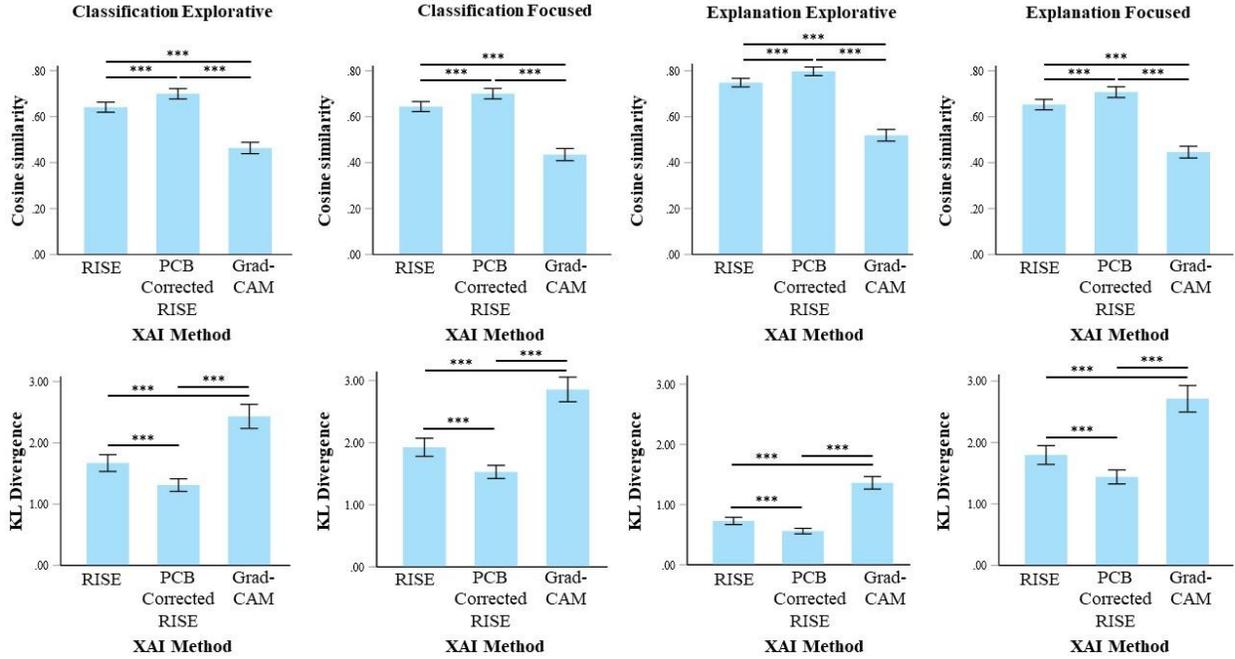

Figure 5. Difference among the three XAI methods (RISE, PCB Corrected RISE, GradCAM) in cosine similarity (row above) and KL divergence (row below) in each of the four task and strategy combinations (error bars: 95% CI; $^*p < .05$, $^{**}p < .01$, $^{***}p < .001$).

## Discussion

Here we examined human attention strategies for image classification and for explaining image classification and compared them with current saliency-map based XAI explanations. Using EMHMM, we discovered focused (focused visual scanning on the foreground object) and explorative (explorative scanning at a broader region) attention strategies in both classification and explanation tasks. Participants did not consistently adopt the same strategies across the two tasks, and they adopted more explorative strategies for explanation than the classification task itself. This result suggested humans adjust their attention strategies according to the task demand (Hsiao, An et al., 2021; Kanan et al., 2015; Chuk, Chan et al., 2017). In addition, in image classification, focused strategies predicted faster response. In contrast, in explanation, focused strategies were associated with explanations that were more diagnostic for inferring image classes, whereas explorative strategies were associated with higher frequency to attend to the image region and explanations rated higher for effectiveness for early category learning. Interestingly, current saliency-based XAI explanations were more similar to human attention strategies in the explanation task, especially the explorative strategies, rather than those during image classification.

In particular, saliency maps generated by perturbation-based XAI methods including PCB corrected RISE and RISE, which highlight input features that lead to output class probability change when being perturbed, had higher similarity to human attention maps than the backpropagation-based XAI method GradCam. This result was consistent with previous research suggesting causal reasoning based on observed regularities as an important feature in human explanations (Einhorn & Hogarth, 1986; Holzinger et al., 2019; Zemla et al., 2017).

The finding that participants used more focused attention strategies for image classification and more explorative strategies for explanation was consistent with our hypothesis that human attention strategies during explanation may cover more relevant features than those during image classification itself. More specifically, image classification may require attention to just sufficient information for making a classification decision (Hsiao & Cottrell, 2008; Smith & Ratcliff, 2004), whereas explanation may require attention to as much relevant information as possible to be comprehensive (Gelman et al., 1998). Consistent with this finding, here we found that in image classification, a more focused attention strategy had faster response speed, suggesting that focusing on identifying critical features of the foreground object is beneficial for classification. In explanation, a more explorative attention strategy was associated with higher frequency to attend to the image region than the textbox region, higher ratings in effectiveness, and more use of visual information in the explanation text, as compared to the focused strategy. This result suggested that explorative scanning for relevant visual features to the object class is beneficial for providing explanations for early category learning. In contrast, explanations with a more focused scanning on the foreground object were rated lower for effectiveness but were associated with higher diagnosticity for inferring the class label and more use of conceptual information in the explanation text. We also found that visual information enhanced effectiveness, consistent with the previous finding that visual information is more important for early category learning (Kloos & Sloutsky, 2008). In contrast, conceptual information made it easier for people who already knew the image classes to guess the class label, consistent with the finding that conceptual information is more dominant in category representations developed later (Fisher & Sloutsky, 2005). Thus, attention strategies revealed participants' preference in information use during explanation, which was in turn associated with explanations that served different purposes. Interestingly, participants' attention strategies were also significantly more explorative when providing explanations for natural than artificial object classes, suggesting higher reliance on visual information, and this result was consistent with previous research reporting that perceptual similarities are more important for the categorization of natural objects than artificial objects (Stibel, 2006). EMHMM allowed us to discover representative attention strategies and quantify individual strategies, leading to these novel findings.

Although saliency-map based XAI methods are designed to highlight features used by AI for performing the task, we found that saliency maps generated by XAI for image classification had higher similarity to human attention strategies during explanation than during the image classification task itself. This finding suggests that the current XAI methods highlight all features that are relevant to AI's classification decision, similar to how humans explain image classification. However, AI's decision processes may be fundamentally different from humans'. More specifically, a fundamental difference between humans and AI is in their attention mechanisms: humans process bits of visual information at a time through a sequence of eye fixations, whereas AI models do not have this visual anatomy constraint and can process all information simultaneously (Hsiao, An, et al., 2022). Thus, human decisions involve accumulation of evidences sequentially (Lee & Cummins, 2004), whereas in AI all relevant information can be processed in parallel (e.g., Raschka et al., 2020). We also found that XAI saliency maps had higher similarities to the explorative than the focused attention strategy during human explanations. As discussed earlier, in humans the explorative strategy was associated with higher reliance on visual information, whereas the focused strategy was associated with higher reliance on conceptual information. Thus, another difference between human and AI image classification is the type of information available for decision making: the AI model under examination is designed to use visual information only; in contrast, human representations for object classes contain both visual and conceptual information (Martin et al., 2018), which can be flexibly and selectively attended to for decision making. Also, in human category learning, both visual exemplars and verbal explanations play an important role: verbal explanations provide crude rules for the category structure, while visual exemplars can be used for finer adjustments based on these rules (Moskvichev et al., 2019). Although some semantic information may be learned from image statistics in deep learning AI models, they do not usually learn explicit verbal rules as humans. Future work may examine ways to incorporate abstract conceptual information in both image classification AI and XAI methods.

Among the three XAI methods examined, we found that the saliency maps generated by perturbation-based methods (RISE and PCB corrected RISE) consistently had higher similarity to human attention maps than those from the backpropagation-based method (Grad-CAM). This result suggested that human attention strategies during explanation may be more similar to the perturbation-based than the backpropagation-based XAI approach. Perturbation-based methods highlight input features that have causal influence on the classification output probability. In contrast, backpropagation-based methods highlight features according to the gradient output class score in a particular input layer. Indeed, human explanations are typically characterized by the emphasis on observable causality (Einhorn & Hogarth, 1986; Holzinger et al., 2019). Humans prefer to use contrastive explanation strategies, often through counterfactuals to illustrate differences between fact and foil outputs, which can be especially effective at establishing causal attribution (Miller, 2019; Miller, 2021). Since perturbation-based methods highlight input features that have causal influence on the given output class relative to other object classes, the resulting saliency map may better match human attention strategy for explanation and potentially more accessible to AI users. Indeed, in human category learning, contrastive explanations with exemplars highlighting features discriminative of different categories have been shown to facilitate learning performance as compared with non-contrastive explanations with within-category exemplars (e.g., Kang & Pashler, 2012; Nosofsky & McDaniel, 2019; Hammer et al., 2009). Similarly, saliency maps from perturbation-based methods may better facilitate user understanding of AI than backpropagation-based methods. Future work may examine this possibility.

Recent research on XAI has proposed to use human attention, such as human manual annotation (Mohseni et al., 2021), as a benchmark for evaluating quality of saliency-map based explanations. Our current results also contribute to this literature by take individual differences into account. More specifically, since individuals differ significantly in attention strategies in explanation, which were associated with different aspects of explanation quality, EMHMM helped discover strategies that suit the explanation needs as a benchmark. Interestingly, the discovered attention strategy for explanations that included more visual information and were rated higher for effectiveness, i.e., the explorative strategy, also had higher similarity to saliency maps generated by XAI methods than the other strategy. Since collecting a large amount of human attention data for benchmarking purposes is often time-consuming, Yang et al. (2022) developed a Human Saliency Imitator model to automatically generate a human attention map given an input image using a deep learning model trained with human attention data with high accuracy (Pearson

Correlation Coefficient = 0.88 on validation). These simulated data can also be used for developing other applications that require human attention data such as human-in-the-loop systems (Gil et al., 2019), demonstrating the importance of simulated data as a new trend in AI/cognitive science research (de Melo et al., 2021).

Our discovery of different explanation strategies from human explainers also suggested that explainees may differ in the type of explanations that is more accessible to them. Indeed, human explainers are shown to use a mixture of visual or conceptual information in their explanations and may adjust their strategy according to explainees' needs (Kaufman & Kirsh, 2022; Strauss & Ziv, 2012). Thus, future XAI development may consider learners' preferences for providing more accessible explanations. Indeed, in order to enhance human-AI interaction, most recent AI research has started to consider the importance of enabling AI to infer humans' mental state, as inspired by an important cognitive capacity in human social interaction, theory of mind (ToM), i.e., the capacity to understand others' behavior by attributing mental states to them (e.g., Akula et al., 2022). This concept applies to XAI development as well, that is, to consider users' backgrounds and preferences when providing explanations, similar to human explainers. For example, it may be beneficial to use more visual information when explaining novel categories or explaining to young children without much category knowledge. In addition, Hammer et al. (2009) discovered that in contrast to older children and adults, young children had difficulties with identifying between-category differences and thus learned better through comparing same-class exemplars. In this case, prototype-based XAI methods, which use the most representative objects of the category as explanation exemplars, may be more suitable. Indeed, humans are shown to prefer prototype-based approaches during early category learning (e.g., Smith & Minda, 1998; Minda & Smith, 2001).

So far we have discussed the similarities and differences between XAI and human explanation strategies. Humans and XAI may also work together to improve explanation quality. This concept of human-in-the-loop system has been adopted in AI design. For example, task-driven human attention can be integrated into AI systems to boost their performance, especially when humans provide better information extraction strategies (Lai et al., 2020; Rong et al., 2021). Future studies may explore how human attention can be incorporated into XAI methods to make XAI's explanations more accessible to humans. For instance, saliency-based XAI methods highlight important regions without providing a logical temporal sequence for users to understand the links among them (Kaufman & Kirsh, 2022). Integrating human attention, which contains temporal information, may help guide users to reach better comprehension. XAI with the ability to infer user strategy may compare it with AI's strategy and inform user when to trust or not to trust AI.

In conclusion, here we showed that human explanation for image classification involves exploring more relevant features than the classification task itself, which only requires sufficient information for decision making. Humans also differed in the use of explorative vs. focused attention strategies during explanation. These strategies were associated with differential reliance on visual and conceptual information in the explanation that served different purposes. The finding that features used by AI as revealed by current saliency-based XAI methods had the highest similarity to the explorative explanation strategy in humans demonstrated a fundamental difference between AI and human: AI could use all relevant information in parallel, whereas human attention involved sequential processing to accumulate evidence. Interestingly, XAI saliency maps that highlight discriminative features informing causality matched better with human attention strategies for explanation, suggesting that establishing causality characterizes human explanation and can potentially make explanations more accessible to AI users. These findings have important implications for developing user-centered XAI methods to enhance human-AI interaction.

## Methods

**Participants**

We recruited 62 participants (52 females[2]), aged 18 to 37 years ($M = 22.5$, $SD = 3.8$) from the University of Hong Kong. They had normal or corrected-to-normal vision. Among the 62 participants, 7 had English, 46 had Chinese, 3 had Korean, 2 had Indonesian, 1 had Hindi, 1 had Sinhalese, 1 had Thai, and 1 had Vietnamese as their first languages. For the participants whose native language was not English, they started to learn English at an average age of 5.2 ($SD = 2.4$). The participants had a mean score of 71.20% ($SD = 12.79\%$) on the Lexical Test for Advanced Learners of English (LexTALE; Lemhöfer & Broersma, 2012), which served as an indicator of their English proficiency. Here we examined the difference between two participant groups using different eye-movement patterns in classification and explanation performance. A power analysis of independent sample t-test based on a similar study comparing eye movement pattern groups on face recognition performance (Chuk, Crookes, et al., 2017; d = 2.18) suggested that a sample size of 52 was sufficient for testing the group effect (d = 0.8, α = .05, β = .2). In addition, we examined whether eye-movement patterns can predict the participants' performance in the two tasks. A power analysis of linear multiple regression indicated that 55 participants were required assuming a medium effect size ($f^2$ = .15, α = .05, β = .2) and one tested predictor (i.e., eye movement pattern). We recruited 62 participants to allow for attrition.

---

[2] There was no gender difference in eye-movement patterns during image classification, $t(59) = 0.47$, $p = .642$, or explanation, $t(60) = 0.43$, $p = .671$. Female and male participants also did not differ in classification performance (accuracy: $t(60) = 0.46$, $p = .645$; RT, $t(53) = 1.09$, $p = .279$) or explanation performance (effectiveness: $t(60) = 0.63$, $p = .529$; diagnosticity: $t(60) = 0.59$, $p = .559$).

**Materials and Apparatus**

The stimuli included 160 images in 20 categories, with 8 images in each category. The 20 image categories contained 9 natural categories, including ant, corn, horse, jellyfish, lemon, lion, mushroom, snail, and zebra, and 11 artificial categories, including broom, cellphone, fountain, harp, laptop, microphone, pizza, shovel, sofa, tennis ball, and umbrella. These image classes were selected from human basic level categories (e.g., Markman & Wisniewski, 1997; Wang et al., 2015) and were also commonly used as output categories of image classification AI models (e.g., Russakovsky et al., 2015).

The 16 images for the categories horse and sofa were obtained from PASCAL VOC (Everingham et al., 2010), while the rest 144 images were from ImageNet (Deng et al., 2009). We have ensured that the images together constituted a representative set, such that images with different levels of foreground object complexity and background saliency were included. All images were resized to fit into a 400 × 520 pixel frame on a blank canvas. Since the original images differed in their aspect ratios, white edges were added to ensure that the images were converted to the same size without any distortion.

The experiment was conducted using E-Prime 3.0 with the extensions for EyeLink (Psychology Software Tools) on a 255 mm × 195 mm laptop with a resolution of 1024 × 768 pixels. Each image spanned 9.68° × 12.32° of visual angle at a viewing distance of 60 cm. The dominant eyes of participants were tracked with an EyeLink Portable Duo eye tracker (SR Research), and a chinrest was used to minimize head movement. A nine-point calibration and validation procedure was performed at the beginning of the classification and explanation task, and re-calibration took place whenever drift correction error was over 1° of visual angle.

**Design and Procedures**

Participants completed two main tasks, including a classification task and an explanation task, four cognitive tasks, and an English proficiency test (LexTALE). Their eye movements were recorded in the two main tasks.

*Classification Task*

In the classification task, the participants were instructed to assign a class label to the 160 images one at a time based on the 20 labels shown at the beginning of the experiment (Figure 6a). Each trial started with a drift check at the center of the screen. After a stable fixation at the center was observed, a fixation cross was shown at the upper left corner of the screen. The image appeared once the participant fixated on the cross for more than 250 ms to ensure that the first fixation on the image was planned by the participants. The participant's fixation was consistently directed to the left and the images were consistently placed on the right to match the reading direction of English (Spalek & Hammad, 2005). After seeing the image, the participants named the class label aloud as quickly as they recognized it, and the image disappeared when their response was detected. The reaction time (RT) was recorded by a microphone through the voice key of the Chronos response box (Psychology Software Tools), and the answer was manually recorded by the experimenter.

*Explanation Task*

In the explanation task, the participants were shown the same 160 images along with the correct labels for each image and were asked to provide an explanation in a textbox about why the specific label should be assigned to the image based on how they classified the same image in the previous task (Figure 6b). They were told to imagine explaining to someone without any prior knowledge of the visual categories, such as a very young child, and to include enough information to help someone learn how to identify the categories. Similar to the classification task, each trial started with a drift check at the center of the screen and a fixation cross directed the participant's fixation to the upper left corner, where the class label appeared. Each trial ended when the participants pressed Enter after they finished typing the explanation.

Before starting the explanation task, the participants were given the following example explanation for an image of an elephant: a long trunk below the eye next to a white pointy object that looks like a tusk; a big triangular ear. They also completed three practice trials with images from three different classes that were not among the 20 classes used in the formal trials. Feedback was given if the explanations did not meet the requirements.

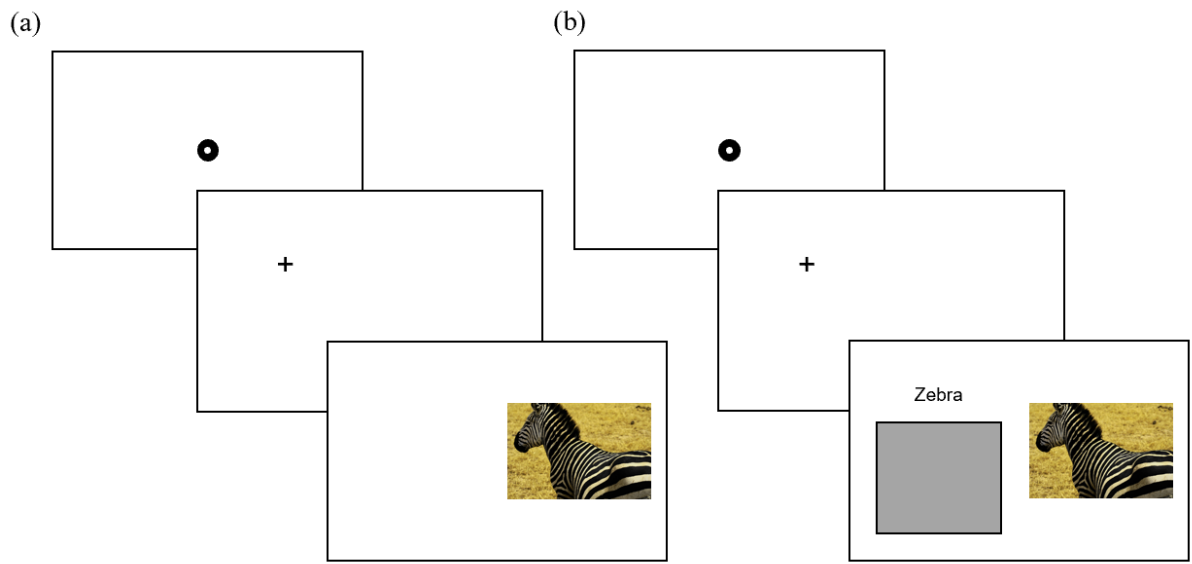

**Figure 6**. Experimental procedure of the (a) classification task and the (b) explanation task.

***Cognitive Tasks and English Proficiency Test***

    **Verbal and Visuospatial Two-Back Tasks**. The two-back tasks (Lau et al., 2010) were used to measure the participants' working memory capacity. In the verbal two-back task, numbers were presented to the participants at the center of the screen one at a time, and the participants judged whether it was the same as the one shown two trials back. In the visuospatial two-back task, different symbols appearing at different locations were presented one at a time, and the participants judged whether they appeared at the same location as the one shown two trials back. They were instructed to disregard the identity of the symbol and to only focus on the location. Each task was conducted in two blocks, and each block included 28 trials. Accuracy and RT were measured for each task.

    **The Flanker Task**. The flanker task (Ridderinkhof et al., 1999) tested the participants' selective attention. In each trial, the participants were presented with five arrows and instructed to judge the direction of the central arrow. The other four arrows, or the flankers, pointed to the same direction as the central arrow in the congruent condition and pointed to the opposite direction in the incongruent condition. In the neutral condition, the flankers would not suggest any directional

response. There were 20 trials for each of the three conditions. Flanker effect in accuracy and RT was measured as (Congruent – Incongruent)/(Congruent + Incongruent).

**Multitasking Test**. The multitasking test assessed the participants' task-switching ability (Stoet et al., 2013). The participants were shown four types of figures differentiated by outer shape (diamond or square) and inner filling (two dots or three dots). Each figure appeared either at the upper section or the lower section of a rectangular box. The participants were asked to judge the shape of the figure if it appeared at the upper section and to judge its filling if it appeared at the lower section. The task included three blocks. The figures always appeared at the top in the first block and always appeared at the bottom in the second block (no-switching tasks). In the third block, the figures appeared randomly either at the top or at the bottom (dual task). Each block contained 32 trials, where each combination of shape and filling (and position, for the third block) repeated for the same number of times. Their task switching ability was measured as the accuracy/RT in the dual task during the third block minus the average accuracy/RT in the two no-switching tasks during the first two blocks.

**Tower of London task**. The participants' executive function and planning ability were evaluated with the Tower of London task (Phillips et al., 2001). On each trial, participants were presented with two figures, each with three balls that were randomly distributed on three sticks. The two figures were respectively the target board and the move board. The participants were instructed to move the balls on the move board to make it look exactly the same as the target board. They were told to use the fewest possible moves and to plan before starting to move. There were 10 trials. Accuracy, averaged number of moves, planning time, execution time, and total time were measured.

**English Proficiency Test**. The LexTALE (Lemhöfer & Broersma, 2012) was used to assess the participants' English lexical knowledge, which can be a potential confounding factor of their performance on the main tasks. This test included 60 trials, where the participants saw a string of letters and judged whether it was an existing English word on each trial.

**Data Analysis**

*Classification Task*

**Task Performance Analysis**. The participants' performance in the classification task was measured by accuracy and average RT. When the participant's response did not match the class label exactly, it was considered correct if it was a synonym (e.g., mobile phone for cellphone) or a closely related word (e.g., tennis for tennis ball).

**Eye-Movement Analysis**. EMHMM (Chuk et al., 2014) with co-clustering (Hsiao, Lan, et al., 2021; see http://visal.cs.cityu.edu.hk/research/emhmm/) was used to model and quantify the participants' eye-movement patterns in the classification task, with both spatial (fixation locations) and temporal (transitions between the locations) dimensions taken into account. Eye movement data from 61 participants on 160 image stimuli were used to perform this analysis.[3] Only fixations on the image area were included in the analysis. Outlier fixations that were more than three standard deviations from the mean fixation location of the specific image on either the vertical or the horizontal dimension were removed. Trials where the participant answered incorrectly were excluded.

Each participant's eye movements in viewing one of the 160 image stimuli were summarized with one HMM, which included person-specific regions of interest (ROIs) and transition probabilities among these ROIs. Thus, each participant would have 160 HMMs if they responded correctly on all of the 160 trials and had no missing trials. A variational Bayesian approach (Coviello et al., 2014) was used to determine the optimal number of ROIs for each individual HMM with a preset range of possible number of ROIs from 1 to 10. Each HMM with a specific number of ROIs was trained for 200 times, and the HMM with the highest log-likelihood within the preset range was chosen.

The co-clustering method was used to cluster the participants into two groups such that participants in the same group had similar eye movement patterns across the stimuli. In other words, the cluster/group assignments are shared across the stimuli, with a representative HMM generated for each group for each stimulus. The number of ROIs of the representative HMMs were set to be the median number of ROIs of the individual HMMs. The co-clustering procedure was repeated for 200 times to select the result with the highest log-likelihood.

The participants' eye-movement patterns were quantified using the A-B scale, which was defined as $(L_A-L_B)/(|L_A|+|L_B|)$, where $L_A$ and $L_B$ represent the log-likelihoods of a participant's

---

[3] One participant was excluded from this analysis due to suspected tracking error. In addition, six participants had inaccurately measured classification RT and were excluded from the subsequent analyses on this variable, but they were still included in the co-clustering analysis because their eye movement data were unaffected. Thus, the effects can be tested with sufficient power provided the finalized sample size (55 valid participants).

eye-movement data being classified as belonging to Pattern Group A and Pattern Group B, respectively (Chan et al., 2018; Liao et al., 2022; Zheng & Hsiao, 2022). A higher A-B scale indicates higher similarity to Pattern Group A and a lower A-B scale indicates higher similarity to Pattern Group B. In addition, data log-likelihood $L_A$ and $L_B$ were used to evaluate whether the two representative patterns differ significantly from each other: If the two groups indeed differed significantly, it was expected that Pattern Group A participants should have significantly higher $L_A$ than $L_B$ and that Pattern Group B participants should have significantly higher $L_B$ than $L_A$ (Chuk et al., 2014; Hsiao, Lan, et al., 2021). Eye movement consistency of the participants were assessed by calculating the entropy for each HMM and summing over all the stimuli (Cover & Thomas, 2006). Higher entropy would indicate less predictability, less consistency, and more randomness in the eye movement patterns.

To examine the relationship between eye movement pattern and classification performance while controlling for English proficiency, we conducted ANCOVA analyses on classification accuracy and RT with pattern group as the independent variable and LexTALE as the covariate variable. We also performed correlation analyses between A-B scale and the two performance measures. In addition, we used hierarchical regression analyses to investigate whether eye movement pattern could predict performance after cognitive abilities and English proficiency were controlled.

*Explanation Task*

**Task Performance Analysis**. The participants' performance on the explanation task was based on the quality of the explanations that they provided. Two performance measures were used: (1) effectiveness for teaching image classification to someone without prior category knowledge as rated by two computer vision experts, and (2) diagnosticity as measured by naïve observers' performance in inferring the image class from the explanation. Typos, misspelled words, misused words, and grammatical errors that may impede understanding were fixed before the evaluations.

To obtain the performance measure on effectiveness, two raters were asked to rate the quality of each explanation on a scale from 1 to 7, where 1 indicated a very poor explanation and 7 indicated a very good explanation. The instructions that the participants received were summarized to them, and it was emphasized that the quality of the explanations should depend on whether it can effectively teach someone without prior knowledge how to classify images based on its visual features or characteristics. For each image, all of the explanations were presented on the same spreadsheet in a random order. The image was included in the first row, which was locked to the top of the spreadsheet so that the image would remain visible as the rater scrolled down to view all of the explanations. Two data scientists with expertise in computer vision were selected as the raters, since they had more experience in processing images in terms of visual features and thus were better at identifying features that are important for classification. The two raters had a good inter-rater reliability in terms of their rating for all participants, Cronbach's alpha = .858. Average rating was used as the measure of effectiveness.

To obtain the performance measure on diagnosticity, the explanations provided by the participants were presented to another group of 124 participants (88 females), aged 18 to 32 years ($M = 20.08$, $SD = 2.03$) without the image and the label. Each participant viewed 160 explanations and were asked to guess the category label for each explanation. Each explanation was evaluated by two participants, and it was ensured that none of the participants saw multiple explanations provided by the same participant for the same category. This group of participants had an average score of 74.26% ($SD = 14.29\%$) on the LexTALE. Responses that matched the original label word or any of its synonyms, hyponyms, or close hypernyms were counted as correct. Percent accuracy was used as the measure of diagnosticity.

**Explanation Text Characteristics Analysis.** The characteristics of the explanation text were quantified by two measures: (1) visual strength to assess reliance on visual information and (2) WordNet similarity to the class label to reflect reliance on conceptual information. The explanations were tokenized and lemmatized using spaCy (Honnibal et al., 2020). The two measures were obtained for each word, and the mean scores were computed for each explanation.

The visual strength measure was retrieved from the Lancaster Sensorimotor Norms (Lynott et al., 2020), which include ratings for how much a word is experienced through different perceptual senses or through actions performed by different body parts. For instance, "black and white stripes" is an explanation with high visual strength. WordNet similarity was calculated with the NLTK interface (Bird et al., 2009) for WordNet (Miller, 1995), which organizes words into sets of synonyms and connects the sets with semantic relations. We used path similarity, which is based on the inverse of the shortest path between two words in the hypernym/hyponym taxonomy, to measure the similarity between each word in the explanations and the label word. For instance, "chair" has high similarity to "sofa." Similarity was calculated only for the nouns since WordNet does not link words with different parts of speech, and the first sense, or meaning, was always used for words with multiple senses.

**Eye-Movement Analysis.** The same EMHMM with co-clustering procedure was used to analyze the 62 participants' eye movements on 160 images during the explanation task. Trials where the participants responded incorrectly on the classification task were excluded, since they were told to provide the explanations based on how they classified the images in the previous task. The procedures and settings were exactly the same as those used for the analysis of the classification task. In particular, this analysis also included only the fixations on the image area. In addition, the gaze preference of the image area and the textbox area were calculated respectively using the average percentage of fixations on the image/textbox area in each trial.

The same analyses as those used for the classification task were conducted to examine the relationship between eye-movement pattern and performance in the explanation task. In addition, correlations between explanation performance, text characteristics, and A-B scale were tested to examine the potential relationship of the use of visual or conceptual information in the explanations with eye-movement patterns and explanation performance.

*Comparison of the Two Tasks*

Following a previous study (Hsiao, An, et al., 2021), the consistency between the participants' eye-movement patterns across the classification and the explanation tasks was examined through examining the correlation between the A-B scales in the two tasks. In addition, we performed another EMHMM with co-clustering with the eye-movement data on the image area from both tasks to discover representative eye movement pattern groups across the two tasks in order directly compare eye movements in the two tasks using A-B scale. More specifically, the analysis included 61 participants' eye movement data in the classification task and 62 participants' eye movement data in the explanation task, resulting in 123 participant-task combinations. One HMM was generated for each participant-task combination and each image stimulus. As participants did not necessarily use the same attention strategy in the two tasks, co-clustering clustered each participant-task combination, instead of each participant, into two pattern groups. Thus, each participant had two A-B scales, one for each task. The other parameters and procedures were the same as those used in the analyses of the two tasks separately.

A Chi-square test was conducted to examine whether the participants' HMMs for a certain task tended to be classified into a certain pattern group. In addition, a 2 × 2 repeated-measures ANOVA was conducted on A-B scale with task and image type (natural vs. artificial) as the independent variables to further examine whether the participants' eye movement patterns differed across the two tasks and to consider the potential effect of image type, since previous research found differences in eye movements when viewing natural and artificial scenes (Momtaz & Daliri, 2016). Moreover, Stibel (2006) found that categorization of natural kinds was based on perceptual similarities, while artifact categorization relied more on rule-based connections between features, suggesting that there can be differences in classification and possibly explanation processes between the two image types.

***Comparison with XAI Saliency Maps***

In this study, we also aimed to compare human participants' attention strategies in the two tasks with the saliency-map based explanations generated by current XAI methods. We chose a pre-trained ResNet-50 from PyTorch (Paszke et al., 2019) as our image classification AI, which is a convolutional neural network that performs image classification with high accuracy. ResNet-50 contained 48 convolutional layers along with an average pooling layer and a max pooling layer. It extracted visual features from the images for classification without any other types of features (e.g., conceptual information) of the classes. For the purpose of this comparison, the participants' fixation data were converted to heatmaps.

One heatmap was plotted for each task and each eye movement pattern group, resulting in four heatmaps for each image. Only the fixations on the image area were used and the pattern group assignment was based on the analyses of the two tasks separately. When generating each heatmap, a zero matrix that matched the size of the image in pixels was initialized and each fixation point was marked as 1 in the matrix according to its x- and y-coordinates. A Gaussian filter with a standard deviation of 21 pixels, which corresponded to 0.5° of visual angle during the experiment, was applied to the matrix for plotting the heatmap.

We selected commonly used XAI methods, including RISE, PCB corrected RISE, and Grad-CAM, for XAI saliency map generation. The XAI methods provided explanations for the output of a pre-trained ResNet-50 from PyTorch (Paszke et al., 2019)[4]. The matrices of the saliency maps were normalized by setting the largest value to 1 and the smallest value to 0, while the other values were adjusted based on the same scale, so that they would be comparable to each other and to the human fixation heatmaps. Two similarity metrics were used to compare the saliency maps with the heatmaps, including cosine similarity and KL divergence. Cosine similarity measures the similarity between two vectors in an inner product space by computing the cosine of the angle between them. The two maps to be compared were first flattened into vectors and their cosine similarity was then calculated using the following formula, where **X** and **Y** represent the vectors for two maps and ‖**X**‖ and ‖**Y**‖ represent the Euclidean norms of the two vectors:

$$\cos(\theta) = \frac{\mathbf{X} \cdot \mathbf{Y}}{\|\mathbf{X}\|\|\mathbf{Y}\|}$$

---

[4] Due to the input constraints of ResNet-50, all the images were adjusted to the size of 217 × 217 pixels before generating the heatmaps and the saliency maps.

KL divergence quantifies the difference between two probability distributions and was computed according to the following formula, where $P$ and $Q^D$ represent the probability distributions corresponding to the two maps and $\epsilon$ represents a very small value:

$$KL(P, Q^D) = \sum_i Q_i^D \log(\epsilon + \frac{Q_i^D}{\epsilon + P_i})$$

For each image, cosine similarity and KL divergence were computed to compare each of the three XAI saliency maps with each of the four human fixation heatmaps. Two $2 \times 2 \times 3 \times 2$ by-items ANOVA analyses were conducted on the similarity metrics with task (classification vs. explanation), attention strategy (explorative vs. focused), and XAI method (RISE vs. PCB corrected RISE vs. Grad-CAM) as the within-item variables and image type (natural vs. artificial) as the between-item variable. Through these analyses, we tested whether XAI saliency maps had higher similarity to human attention during classification or explanation, and examined which XAI method provided saliency-map based explanations that had the highest similarity to human attention during explanation.

## Acknowledgements

This study was supported by Huawei. The eye tracker used was supported by RGC of Hong Kong (No. C7129-20G to Hsiao). We thank Van Kie Liew for data processing, Yundi Yang for data collection and preprocessing, Yunke Chen for data preprocessing, Yannie Lim for data collection, Luyu Qiu and Jindi Zhang for their advice.

## Author contribution statements

JH conceptualized the study, acquired the funding, designed the experiment, interpreted the data, contributed to the original draft and substantively revised it. RQ and YZ designed the experiment, developed the experiment programs, collected, preprocessed, analyzed and interpreted the data, and wrote the original draft under the supervision of JH. YY contributed to the conceptualization, experiment design, and data visualization. CC contributed to the conceptualization and revised the manuscript. All authors read and approved the final manuscript.

## Competing interest statements

The authors declare no competing financial and/or non-financial interests.